\documentclass[12pt,a4paper,dvips]{article}
\usepackage{a4p}
\usepackage{cite,mcite}
\usepackage{graphicx}
\usepackage{physics }
\usepackage{l3_title,ifthen,Lep}
\journalname{Phys. Lett. B}
\date{January 06, 2000}
\preprint{2000-005}
\Lep{2}
\graphicspath{{./}}
\newlength{\capindent}
\newlength{\capwidth}
\newlength{\figwidth}
\newcommand{\icaption}[2][!*!,!]{\hspace*{\capindent}%
  \begin{minipage}{\capwidth}
    \ifthenelse{\equal{#1}{!*!,!}}%
    {\caption{#2}}%
    {\caption[#1]{#2}}
  \end{minipage}}
\newcommand{\aqsq}[1]{\ensuremath{\alpha(#1)}}
\newcommand{\iaqsq}[1]{\ensuremath{\alpha^{-1}(#1)}}

\begin{document}
\begin{titlepage}
\title{Measurement of the Running of the Fine-Structure Constant}
\author{The L3 Collaboration}

\begin{abstract}
  Small-angle Bhabha scattering data recorded at the Z resonance and large-angle
  Bhabha scattering data recorded at $\sqrt{s} = 189 \GeV$ by the L3 detector at
  LEP are used to measure the running of the effective fine-structure constant for
  spacelike momentum transfers. The results are
  \begin{eqnarray*}
    \iaqsq{-2.1 \GeV^{2}} - \iaqsq{-6.25 \GeV^{2}}   & = & 0.78 \pm 0.26 \\
    \iaqsq{-12.25 \GeV^{2}} - \iaqsq{-3434 \GeV^{2}} & = & 3.80 \pm 1.29,\\
  \end{eqnarray*}
  in agreement with theoretical predictions.
\end{abstract}
\submitted
\end{titlepage}

%%%%%%%%%%%%%%%%%%%%%%%%%%%%%%%%%%%%%%%%%%%%%%%%%%%%%%%%%%%%%%%%%%%%%%%%%%%%%%%
% Introduction
%%%%%%%%%%%%%%%%%%%%%%%%%%%%%%%%%%%%%%%%%%%%%%%%%%%%%%%%%%%%%%%%%%%%%%%%%%%%%%%
\section*{Introduction}

At zero momentum transfer, the QED~\cite{ref:QED} fine structure constant
\aqsq{0}\ is very accurately known from the measurement of the anomalous magnetic
moment of the electron and from solid-state physics
measurements~\cite{ref:codata}:
\begin{displaymath}
  \iaqsq{0} = 137.035\, 999\, 76\; (50).
\end{displaymath}
In QED, vacuum polarization corrections
to processes involving the exchange of virtual photons result in a $Q^{2}$
dependence, or running, of the \emph{effective} fine-structure constant,
\aqsq{Q^{2}}. This $Q^{2}$ dependence is usually parametrised~\cite{ref:running}
as
\begin{equation}
  \aqsq{Q^{2}} = \frac{\aqsq{0}}{1-\Delta\alpha(Q^{2})}.
  \label{eq:aqsq}
\end{equation}
Whereas the leptonic contributions to $\Delta\alpha(Q^{2})$ can be accurately
calculated, 
due to non-perturbative QCD effects the vacuum polarization contributions from
quark loops cannot be calculated exactly. Therefore, dispersion integral
techniques~\cite{ref:amz1,ref:burkhardt,ref:steinhauser} are used to estimate
these contributions from the measured cross sections for the process
$\epem\rightarrow$ hadrons at low centre-of-mass energies, yielding at $Q^{2} =
\MZ^{2}$: $\iaqsq{\MZ^{2}} = 128.886 \pm 0.090$.
Similar evaluations have been made in Reference~\cite{ref:swartz} and,
under additional theoretical assumptions, in
References~\cite{ref:davier,ref:adler}.

Processes involving photon exchange at non-zero momentum transfer $Q^{2}$ yield
amplitudes proportional to \aqsq{Q^{2}}. At \epem\ colliders this implies that
the measurement of the relative rates of processes involving different $Q^{2}$
values gives access to the running of \aqsq{Q^{2}}\ between those $Q^{2}$
values. This was first exploited by
the TOPAZ Collaboration~\cite{ref:topaz}, which derived a value of the fine
structure constant at a scale of $Q^{2} = 3338 \GeV^{2}$ from the ratio of the
muon pair annihilation cross section to the two-photon-induced muon pair cross
section.

In regions where Bhabha scattering is dominated by $t$-channel photon exchange,
this process allows the study of \aqsq{Q^{2}}\ in the spacelike region,
$Q^{2} = t = -s(1-\cos\theta)/2 < 0$, where $\theta$ is the angle of the outgoing
$\e^{-}$ with respect to the $\e^{-}$ beam direction. The VENUS
Collaboration has recently interpreted large-angle Bhabha
scattering measurements in terms of the evolution of \aqsq{Q^{2}}\ in the range
$100 \GeV^{2} < -Q^{2} < 2916 \GeV^{2}$~\cite{ref:venus}.

This article interprets measurements of Bhabha scattering by the L3 experiment at
LEP in terms of the evolution of \aqsq{Q^{2}}, using two complementary
measurements: first, an analysis of the small-angle Bhabha scattering data in the
polar angular range 32 mrad $<\theta <$ 54 mrad, collected at centre-of-mass
energies around the Z mass and used for the high-precision luminosity measurement
in the years 1993-1995; second, the measurement of the Bhabha scattering cross
section in the polar angular range $20^{\circ} < \theta < 36^{\circ}$ performed at
$\sqrt{s} = 188.7 \GeV$ in 1998. They are used to study the running of
\aqsq{Q^{2}}\ in the range $2.1 \GeV^{2} < -Q^{2} < 6.25 \GeV^{2}$ and 
$12.25 \GeV^{2} < -Q^{2} < 3434 \GeV^{2}$, respectively.

%%%%%%%%%%%%%%%%%%%%%%%%%%%%%%%%%%%%%%%%%%%%%%%%%%%%%%%%%%%%%%%%%%%%%%%%%%%%%%%
% Small-angle Bhabha scattering
%%%%%%%%%%%%%%%%%%%%%%%%%%%%%%%%%%%%%%%%%%%%%%%%%%%%%%%%%%%%%%%%%%%%%%%%%%%%%%%
\section*{Small-angle Bhabha Scattering}

\subsection*{Data Analysis}

The analysis of the small-angle Bhabha scattering data follows closely that of
the luminosity measurement~\cite{L3-SLUM}, and makes use of the same detectors:
\begin{itemize}
 \item two small-angle calorimeters consisting of BGO crystals, providing a
  precise energy measurement for electromagnetic showers. The highest-energy
  cluster reconstructed in each calorimeter is retained for analysis. The energy
  of one cluster must exceed $0.8 E_{\mathrm{beam}}$; the energy of the cluster in
  the other calorimeter must be greater than $0.4 E_{\mathrm{beam}}$. To avoid
  edge effects their reconstructed polar angles must be well contained in the
  calorimeter, $28\;\mbox{mrad} < \theta < 65$ mrad;
 \item a silicon strip detector, consisting of two layers of $r$-measuring
  strips and one layer of $\phi$-measuring strips, installed in front of each BGO
  calorimeter. This detector allows a precise definition of the fiducial
  volume. The coordinates of the cluster reconstructed in the calorimeter are
  projected onto each $r$-measuring silicon detector plane in turn, and a
  matching window (5 mm for $r$-measuring strips and $2.5^{\circ}$ for
  $\phi$-measuring strips) is used to search for corresponding hit strips. If
  found, their coordinates are used. Otherwise the BGO coordinates are retained.
\end{itemize}
Angular cuts are applied on the coordinate measurements for each $r$-measuring
plane in turn.
The fiducial volume is the same as that used for the luminosity measurement (32
mrad $<\theta <$ 54 mrad), divided into four polar angular bins, with boundaries
at 32, 35, 40, 46, and 54 mrad.
For each event, each of the four coordinate measurements (two on each side) is
entered in the corresponding angular bin with a weight of 0.25.

The silicon coordinate reconstruction is sensitive to malfunctioning silicon
strips. Therefore, only those data are used for which the silicon detector was
fully functional, resulting in a sample of $6.7 \cdot 10^{6}$ Bhabha events
corresponding to an integrated luminosity of 98.8~\pb. In
addition, the polar angular bins are chosen to have boundaries sufficiently far
away from malfunctioning strips (8 in 1993, 9 in 1994, and 11 in 1995) and from
the edges of the flare in the beam pipe (42.5 and 50 mrad) on the outgoing
$\e^{-}$ side. The data are grouped into 7 sets, according to the three years
and the different centre-of-mass energies.

Detector effects are corrected for, bin by bin, using a sample of $2.6 \cdot
10^{6}$ fully simulated~\cite{ref:geant} and reconstructed
BHLUMI~\cite{ref:bhlumi} Monte Carlo events. The correction factors differ from 1
by at most $0.4\%$ and have statistical uncertainties of about $4\cdot 10^{-4}$.

The data are compared with predictions from the BHLUMI Monte Carlo program. This
program is modified to allow the running of \aqsq{Q^{2}}\ in the spacelike
region to be different from the nominal one of Ref.~\cite{ref:amz1}. The running
in the timelike region is left unchanged. This is relevant for the interference
term between the dominant $t$-channel photon-exchange and the $s$-channel
Z-exchange diagrams, which contributes up to 0.15\% to the Bhabha cross section in
the luminosity monitor fiducial volume at centre-of-mass energies around the Z
resonance. The predictions are calculated for each data set separately.

Since the small-angle Bhabha scattering data are used for the absolute
normalisation of all processes, only the shape of the polar angular distribution
can be used in this study. Therefore, the compatibility of the data with any
given parametrisation, $p$, of the running of \aqsq{Q^{2}}\ is evaluated in terms
of a likelihood formula according to multinomial statistics:
\begin{equation}
  \label{eq:likelihood}
  -\ln\mathcal{L} = -\sum_i \ln
   P(n_{1,i},\ldots,n_{4,i};f_{1,i}^{p},\ldots,f_{4,i}^{p}),
\end{equation}
where
\begin{displaymath}
  f_{j,i}(p) = \int_{\theta_{ji,\min}}^{\theta_{ji,\max}}
  \frac{d\sigma^{p}}{d\theta} d\theta /
  \int_{\theta_{\min}}^{\theta_{\max}} \frac{d\sigma^{p}}{d\theta} d\theta
\end{displaymath}
is the fractional cross section predicted in bin $j$ ($= 1\ldots4$) for data set $i$
($= 1\ldots7$) for parametrisation $p$, $n_{j,i}$ is the observed event weight in
each angular bin, and $P$ is the multinomial probability distribution.

%%%%%%%%%%%%%%%%%%%%%%%%%%%%%%%%%%%%%%%%%%%%%%%%%%%%%%%%%%%%%%%%%%%%%%%%%%%%%%%
% Results
%%%%%%%%%%%%%%%%%%%%%%%%%%%%%%%%%%%%%%%%%%%%%%%%%%%%%%%%%%%%%%%%%%%%%%%%%%%%%%%
\subsection*{Results}

\begin{figure}[p]
  \begin{center}
    \includegraphics[width=\figwidth]{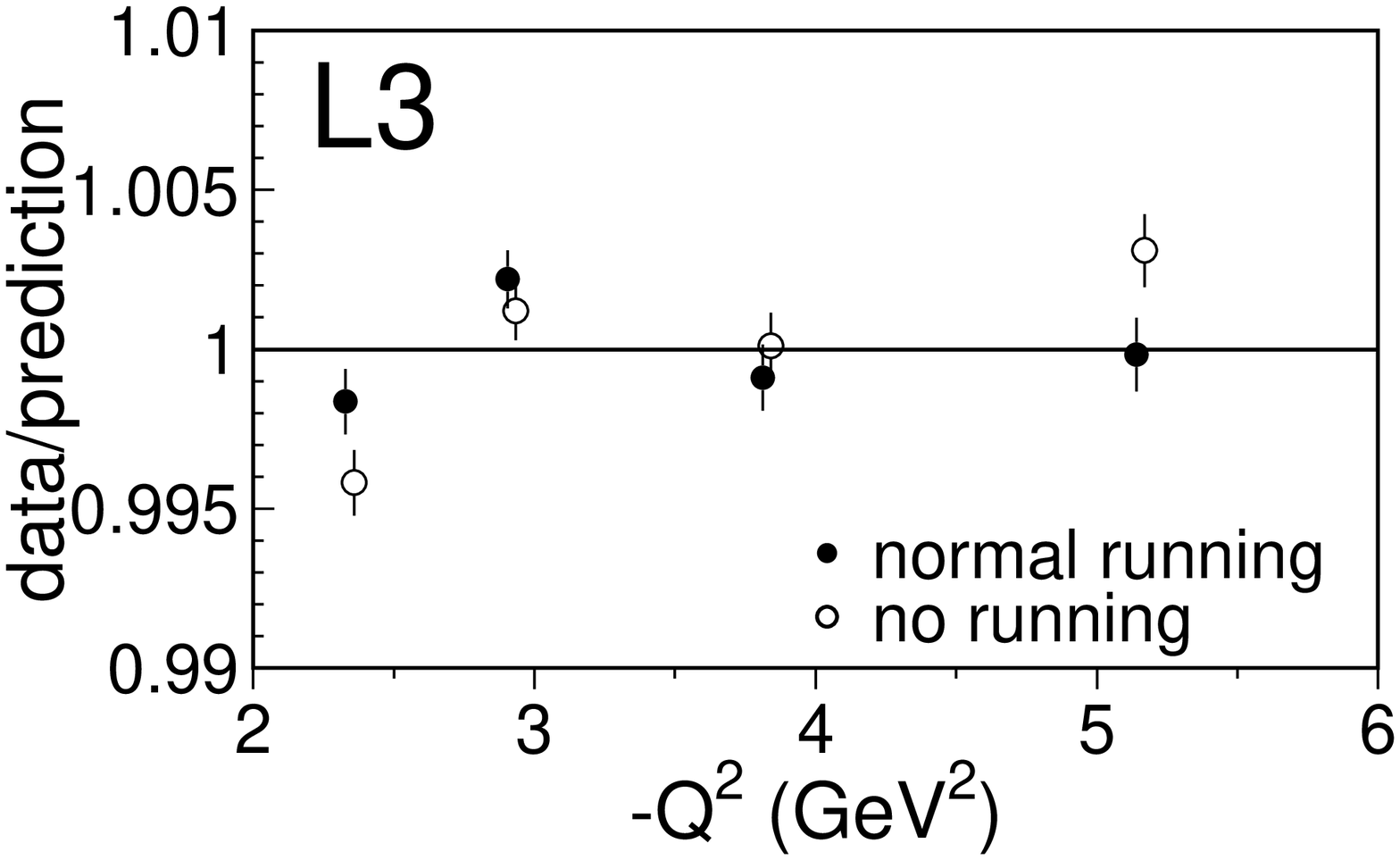}
    \icaption{Ratio of the measured and expected fractional event weights in
      each polar angular bin for the standard theoretical predictions (solid
      circles) and the assumption of no running (open circles), for the
      data collected in 1994.
      \label{fig:simple}}
  \end{center}
\end{figure}

The data are first compared with two hypotheses: the theoretical
prediction~\cite{ref:amz1} (referred to as `normal running') and the
prediction for the case of no running of \aqsq{Q^{2}} between $Q^{2}$ and the
lowest scale accessible in the analysis, $Q_{0}^{2} = -2.1
\GeV^{2}$. Figure~\ref{fig:simple} shows, for the 1994 data which have the
highest statistics, the ratio of the measured and expected fractional event
weights for the two hypotheses.
The likelihood difference between the two hypotheses is
$-(\ln\mathcal{L}(\mbox{no running})-\ln\mathcal{L}(\mbox{normal running})) =
10.24$ for the complete data set, excluding the hypothesis of no running by
$4.5\sigma$.

The data are then compared with a parametrisation in which a term, linear in
$Q^{2}$ and with a slope, $S$, to be fitted, is added to the normal
running. The $Q^{2}$ dependence of Equation~\ref{eq:aqsq} then becomes
\begin{equation}
  \aqsq{Q^{2}} = \frac{\aqsq{0}}{1 - \Delta\alpha(Q^{2}) - S\cdot
  (Q^{2}-Q_{0}^{2})},
  \label{eq:aqsq-lin}
\end{equation}
with $Q_{0}^{2}$ as given above. The fit yields the result $S = (-3.6 \pm 2.7)
\cdot 10^{-4} \GeV^{-2}$, where the error is statistical only.

The following sources of systematic uncertainties on $S$ have been studied:
\begin{itemize}
 \item Statistical uncertainties on the detector correction factors: their
  effect is studied by generating a large number of random number sequences in
  which each correction factor is changed according to a Gaussian
  distribution having a width equal to its statistical uncertainty. Since the same
  Monte Carlo sample is always used for the corrections, the changes in the 
  correction factors are assumed to be fully correlated between the same bins of
  different data sets. Each time the corrections are changed, and the spread in
  the fitted values of $S$ is found to be $1.5 \cdot 10^{-4} \GeV^{-2}$;
 \item Statistical uncertainties on the theoretical predictions: their effect is
  studied in the same way. In this case, the statistical uncertainties on the
  cross sections are assumed to be fully correlated between different choices of
  the parametrisation of \aqsq{Q^{2}}, but to be uncorrelated between different
  data sets. This yields an uncertainty of $0.5 \cdot 10^{-4} \GeV^{-2}$.
 \item The size of the matching window used for the coordinate reconstruction:
  this is studied by changing it to 4 mm and $2^{\circ}$ for the $r$- and
  $\phi$-measuring strips, respectively, and the difference in the fit result
  ($0.8 \cdot 10^{-4} \GeV^{-2}$) is assigned as a systematic uncertainty.
 \item The consistency between the results obtained using the individual
  $r$-measuring silicon layers: instead of combining the results from the four
  $r$-measuring layers, the analysis can also be performed for each side or layer
  separately. The fitted values of $S$ are $-6.7 \cdot
  10^{-4}$ using the layers on the outgoing $\e^{+}$ ($-z$) side, and $-0.5 \cdot
  10^{-4}$ using the layers on the outgoing $\e^{-}$ ($+z$) side. The results
  obtained using each of the two layers on either side are in excellent agreement
  with each other.

  From the ratio of the (corrected) fractions in each bin for the $-z$ and $+z$
  sides, the difference is seen to be due to a rather poor
  overall agreement between the observed fractions for the two sides. The
  cause for this poor agreement is an incomplete simulation of the beampipe
  material traversed at small angles by the $\e^{\pm}$ before entering the silicon 
  detector. The uncertainty assigned to this source is half the difference between
  the results for the two sides, $3.1\cdot 10^{-4}$. This dominates the total
  measurement uncertainty.

  As a cross-check on the material effects, the last two bins are merged into a
  single bin which covers the complete region of increased material in the flare
  of the beam pipe. An alternative is to take out from the analysis the first
  bin, which represents the most material for the layers on the other side. In
  both cases the results are stable within the systematic uncertainty assigned.
  
  Finally, the effect of the choice of polar angular bins is studied by
  performing the fits for several choices of binning. Variations of the size of
  the systematic uncertainty assigned are found, which are almost entirely due
  to variations observed on the $+z$ side. Their cause is the same as that of the
  poor overall agreement, and no additional systematic uncertainty is assigned.
 \item The uncertainty on the vertex position: this affects the two sides in
  opposite ways. Its effect has been estimated by artificially moving the
  vertex position, and is found to be negligible.
\end{itemize}
The largest contribution (0.04\%) to the theoretical systematic
uncertainty on the luminosity measurement at centre-of-mass energies around the Z
resonance~\cite{bhlumi-new} comes from the theoretical uncertainty on the
value of \aqsq{Q^{2}}\ in the small-angle Bhabha scattering region, and hence
should not be considered in this analysis. Theoretical uncertainties on other
contributions can affect the angular dependence only slightly. For example, the
uncertainty due to missing $\mathcal{O}(\alpha^{3}L^{3})$ (where $L$ represents
the `large logarithm' $\ln(|t|/m_{\e})$) contributions to the uncertainty on $S$
is estimated to be about $4\cdot 10^{-6}$, \emph{i.e.} negligible compared to the
experimental systematics.

The final result of the measurement of $S$ is thus 
\begin{equation}
  S = (-3.6 \pm 2.7 (\mbox{stat.}) \pm 3.5 (\mbox{syst.})) \cdot 10^{-4} \GeV^{-2}.
\end{equation}

%%%%%%%%%%%%%%%%%%%%%%%%%%%%%%%%%%%%%%%%%%%%%%%%%%%%%%%%%%%%%%%%%%%%%%%%%%%%%%%
% Large-angle Bhabha scattering
%%%%%%%%%%%%%%%%%%%%%%%%%%%%%%%%%%%%%%%%%%%%%%%%%%%%%%%%%%%%%%%%%%%%%%%%%%%%%%%
\section*{Large-angle Bhabha Scattering}

\subsection*{Data Analysis}

The subdetectors used for the measurement of large-angle Bhabha scattering are
the BGO electromagnetic calorimeter and the central tracker, described in detail
in Reference~\cite{l3-00}. The data were collected in 1998 at $\sqrt{s} = 188.7
\GeV$ and correspond to an integrated luminosity of 175.9~\pb.
Electrons are identified as energy deposits of at least 0.5~\GeV\ in the
calorimeter, with an electromagnetic transverse shower shape, and at least six
associated track hits within a three degree azimuthal angular range. At least one
electron must be observed within the fiducial volume $20^{\circ} < \theta <
36^{\circ}$, and one within $144^{\circ} < \theta < 160^{\circ}$. At most six 
electromagnetic clusters are admitted in the electromagnetic calorimeter. The
highest-energy electron must have an energy exceeding $0.5 E_{\mathrm{beam}}$.
In this angular range, $t$-channel photon exchange gives the dominant contribution
to the cross section, with $s-t$ interference and $s$-channel exchange
contributing only -7\% and 0.3\%, respectively.

The signal efficiency is estimated using $2.5\cdot 10^{6}$ fully simulated and
reconstructed BHWIDE~\cite{BHWIDE} Monte Carlo events generated in the angular
range $5^{\circ} < \theta < 175^{\circ}$. The background, consisting mainly of 
$\epem\rightarrow\tautau(\gamma)$ events, is small, 0.12\%, and its uncertainty
has a negligible effect on the cross section measurement.

Given the large $Q^{2}$ range covered by this analysis, the parametrisation of
Equation~\ref{eq:aqsq-lin} is not appropriate. 
Deviations from the normal running of \aqsq{Q^{2}}\ are parametrised, for any
$Q^{2}$, using the following modification of Equation~\ref{eq:aqsq}:
\begin{equation}
  \label{eq:aqsqmod}
  \aqsq{Q^{2}} = \frac{\aqsq{0}}{1-C\cdot\Delta\alpha(Q^{2})},
\end{equation}
where $\Delta\alpha(Q^{2})$ is calculated according to 
Reference~\cite{ref:amz1}.
The evolution of \aqsq{Q^{2}}\ is then determined from the ratio of measured to
predicted cross sections, $\sigma_{\mathrm{meas}}/\sigma_{\mathrm{pred}}(C)$,
where $\sigma_{\mathrm{pred}}(C)$ is the predicted cross section for a given
value of $C$. Since Equation~\ref{eq:aqsqmod} also modifies the value of
\aqsq{Q^{2}}\ in the $Q^{2}$ range used for the luminosity measurement, also the
measured luminosity and hence the measured cross section is changed for $C \neq
1$. This is accounted for by the use of the ratio
\begin{equation}
  \label{eq:ratioc}
  R(C) = \frac{\sigma_{\mathrm{meas}}}{\sigma_{\mathrm{pred}}(C)} \cdot
         \frac{\sigma_{\mathrm{lumi}}(C)}{\sigma_{\mathrm{lumi}}(C=1)},
\end{equation}
where $\sigma_{\mathrm{lumi}}(C)$ is the small-angle Bhabha scattering cross
section in the fiducial volume used for the luminosity measurement. $C$ is then
determined by equating $R(C) = 1$. The cross section predictions are obtained
using the BHWIDE and BHLUMI Monte Carlo programs for the large-angle and
small-angle Bhabha scattering processes, respectively, modified to parametrise the
evolution of \aqsq{Q^{2}}\ according to Equation~\ref{eq:aqsqmod}.

%%%%%%%%%%%%%%%%%%%%%%%%%%%%%%%%%%%%%%%%%%%%%%%%%%%%%%%%%%%%%%%%%%%%%%%%%%%%%%%
% Results
%%%%%%%%%%%%%%%%%%%%%%%%%%%%%%%%%%%%%%%%%%%%%%%%%%%%%%%%%%%%%%%%%%%%%%%%%%%%%%%
\subsection*{Results}

The number of selected events is 23940. The selection efficiency is estimated to
be $(93.38 \pm 0.08)$\%, where the error is due to limited Monte Carlo
statistics. The measured cross section is $\sigma_{\mathrm{meas}} = 145.6
\pm 0.9 \;\mathrm{pb}$.

Systematic uncertainties on the cross section measurement are due to:
\begin{itemize}
 \item The description of the energy response: Figure~\ref{fig:e1} shows the
  energy of the highest-energy electron, normalised to $E_{\mathrm{beam}}$. The
  low-energy tail of the distribution is not very well described by the Monte
  Carlo simulation. Its effect on the selection efficiency is estimated by
  adjusting the Monte Carlo distribution to obtain a better agreement, and is
  found to be negligible.
  \begin{figure}[p]
    \begin{center}
      \includegraphics[width=0.85\textwidth]{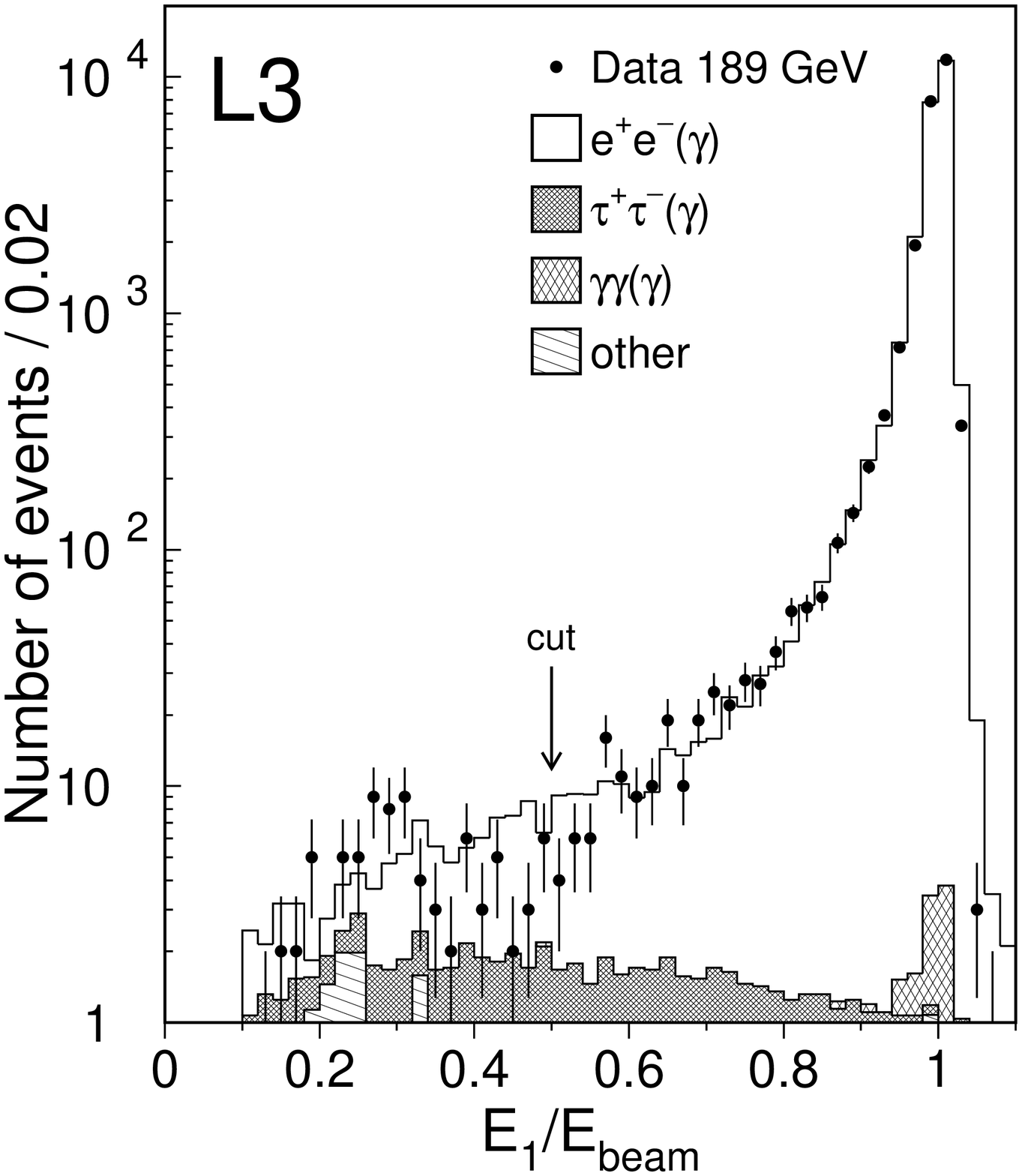}
      \icaption{Distribution of the highest electron energy normalised to the beam
        energy. The arrow shows the position of the applied cut.
        \label{fig:e1}}
    \end{center}
  \end{figure}
 \item The definition of the fiducial volume: this is estimated by varying the
  angular cuts and results in a 0.4\% uncertainty.
 \item The cut on the number of track hits: a comparison between data and Monte
  Carlo events has been used to adjust the Monte-Carlo description of the
  single-hit efficiency of the inner tracker wires. Varying the cut on the
  number of hits changes the efficiency by 0.3\%.
\end{itemize}
Summing the individual contributions in quadrature, a total systematic
uncertainty of 0.5\% is obtained. The result of the cross section measurement is
thus
\begin{equation}
  \label{eq:sigmeas}
  \sigma_{\mathrm{meas}} = 145.6 \pm 0.9 (\mbox{stat.}) \pm 0.8
  (\mbox{syst.})\; \mathrm{pb}.
\end{equation}

The measured cross section agrees well with the theoretical prediction of
$\sigma_{\mathrm{pred}}(C=1) = 145.9 \;\mathrm{pb}$. Using
Equation~\ref{eq:ratioc} and assuming a theoretical uncertainty on
$\sigma_{\mathrm{pred}}(C=1)$ of 1.5\%~\cite{Placzek}, the value of $C$ obtained
is 
\begin{displaymath}
  C = 0.97 \pm 0.12 (\mbox{stat.}) \pm 0.10 (\mbox{syst.}) \pm 0.29 (\mbox{theory}).
\end{displaymath}

\section*{Interpretation of Results}

The results can be interpreted as a measurement of the evolution of
\iaqsq{Q^{2}}\ in the spacelike region between the momentum transfer scales
relevant for the analyses. In the small-angle Bhabha scattering analysis these are
taken to be the lowest and highest $Q^{2}$ values accessible, $-2.1 \GeV^{2}$ and
$-6.25 \GeV^{2}$. The difference is found to be
\begin{displaymath}
  \iaqsq{-2.1 \GeV^{2}} - \iaqsq{-6.25 \GeV^{2}} = 0.78 \pm 0.26,
\end{displaymath}
where the error reflects the total experimental uncertainty.

In the large-angle Bhabha scattering analysis the relevant momentum transfer
scales are taken to be the average $Q^{2}$ value used for the luminosity
measurement, $-12.25 \GeV^{2}$, and the average $Q^{2}$ value for the $t$-channel
contribution to the large-angle Bhabha scattering cross section, $-3434 \GeV^{2}$.
The result is
\begin{displaymath}
  \iaqsq{-12.25 \GeV^{2}} - \iaqsq{-3434 \GeV^{2}} = 3.80 \pm 0.61
  (\mbox{expt.}) \pm 1.14 (\mbox{theory}).
\end{displaymath}

\begin{figure}[p]
  \begin{center}
    \includegraphics[width=\figwidth]{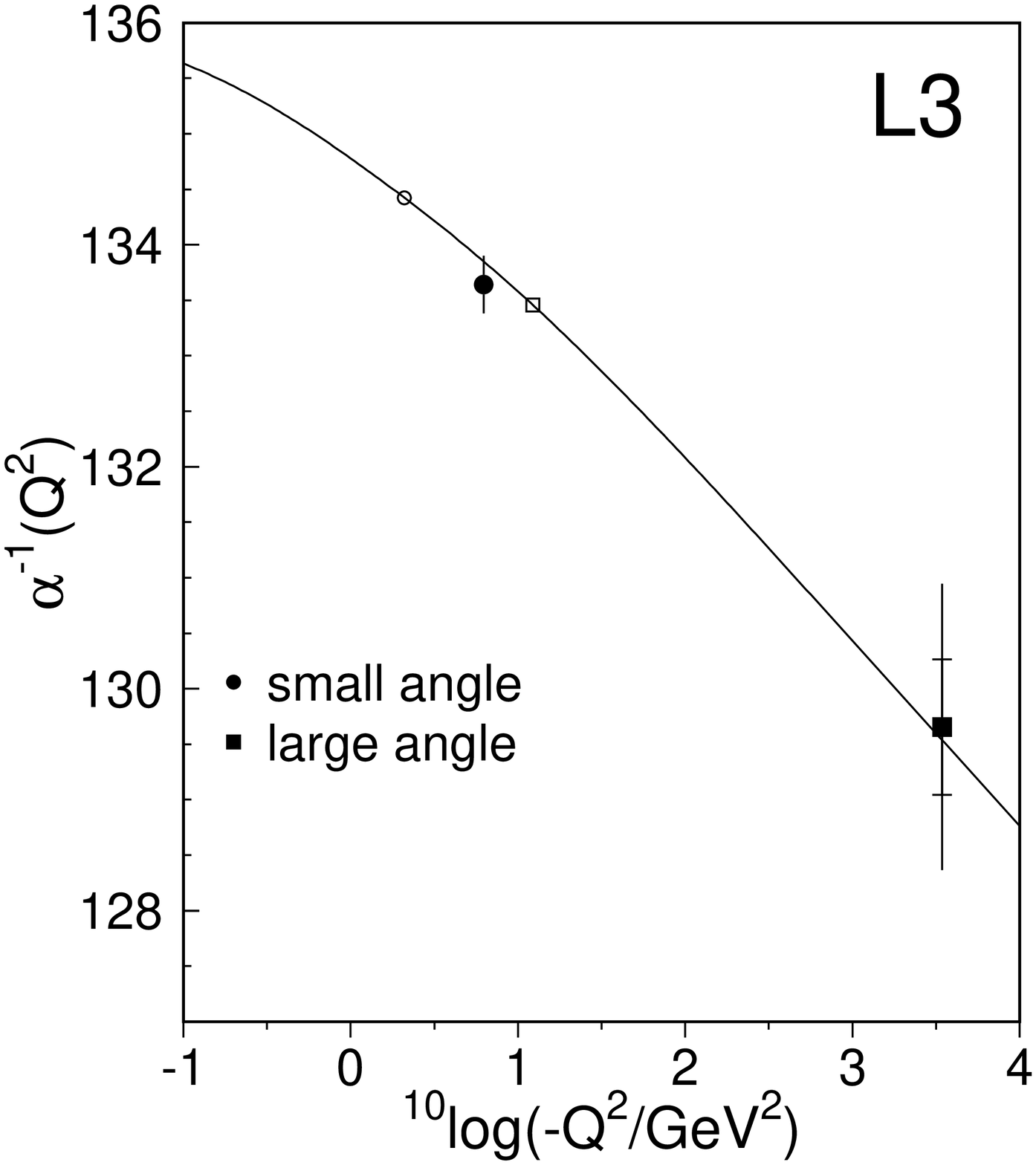}
    \icaption{Measurements of \iaqsq{Q^{2}} for $Q^{2} < 0$.
      The results of the small-angle and large-angle Bhabha scattering
      measurements described in this article are shown as a solid circle and
      square, respectively. The corresponding reference $Q^{2}$ values at which
      the value of \iaqsq{Q^{2}} is fixed to its expectation are shown as open
      symbols. The error bar on the large-angle point indicates the experimental
      and the total uncertainty.
      \label{fig:running}}
  \end{center}
\end{figure}

The results are displayed in Figure~\ref{fig:running}. For the purpose of the
figure, for both measurements the value of \aqsq{Q^{2}}\ is fixed to its
expectation at the lower momentum scale involved in the analysis. 

In conclusion, Bhabha scattering at LEP has been used to study the running of the
fine-structure constant, \aqsq{Q^{2}}, in the spacelike momentum transfer regions
$2.1 \GeV^{2} < -Q^{2} < 6.25 \GeV^{2}$ and $12.25 \GeV^{2} < -Q^{2} < 3434
\GeV^{2}$. The data clearly establish a nonzero running as predicted by QED.

\section*{Acknowledgements}

We are grateful for useful discussions with F.~Jegerlehner.
We wish to express our gratitude to the CERN accelerator divisions for the
excellent performance of the LEP machine. 
We acknowledge the contributions of the engineers and technicians who have
participated in the construction and maintenance of this experiment.  

%
%%%%%%%%%%%%%%%%%%%%%%%%%%%%%%%%%%%%%%%%%%%%%%%%%%%%%%%%%%%%%%%%%%%%%%%%%%%%%%%
% Bibliography
%%%%%%%%%%%%%%%%%%%%%%%%%%%%%%%%%%%%%%%%%%%%%%%%%%%%%%%%%%%%%%%%%%%%%%%%%%%%%%

%%%%%%%%%%%%%%%%%%%%%%%%%%%%%%%%%%%%%%%%%%%%%%%%%%%%%%%%%%%%%%%%%%%%%%%%%%%%%%
% Author List
%%%%%%%%%%%%%%%%%%%%%%%%%%%%%%%%%%%%%%%%%%%%%%%%%%%%%%%%%%%%%%%%%%%%%%%%%%%%%%
\newpage
\typeout{   }     
\typeout{Using author list for paper 197 ONLY }
\typeout{Using author list for paper 197 ONLY }
\typeout{Using author list for paper 197 ONLY }
\typeout{Using author list for paper 197 ONLY }
\typeout{Using author list for paper 197 ONLY }
\typeout{Using author list for paper 197 ONLY }
\typeout{$Modified: Thu Jan 27 16:09:16 2000 by clare $}
\typeout{!!!!  This should only be used with document option a4p!!!!}
\typeout{   }
%
%
%
%  L A T E X  version!!
%
%
% Make sure that the Lep package has been used!
%\input{Lep.sty}%
%
%\ifx\LepCalled\undefined%
%\typeout{     }%
%\typeout{!!!!!!!!!!!!!!!!!!!!!!!!!!!!!!!!!!!!!!!!!!!!!!!!!!!!!!!!!!!}%
%\typeout{Yikes.  You haven't used the Lep package!}%
%\typeout{Please put \protect\usepackage\protect{Lep\protect} in your preamble,
%         followed by}%
%\typeout{\protect\Lep\protect{1\protect} or \protect\Lep\protect{2\protect}}%
%\typeout{     }%
%\typeout{For now you will get a Lep phase 2 authorlist (may not be right!).}%
%\typeout{!!!!!!!!!!!!!!!!!!!!!!!!!!!!!!!!!!!!!!!!!!!!!!!!!!!!!!!!!!!}%
%\typeout{     }%
%\Lep{2}\fi%

\newcount\tutecount  \tutecount=0
\def\tutenum#1{\global\advance\tutecount by 1 \xdef#1{\the\tutecount}}
\def\tute#1{$^{#1}$}
\tutenum\aachen            % 1
\tutenum\nikhef            % 2
\tutenum\mich              % 3
\tutenum\lapp              % 4
\tutenum\basel             % 5
\tutenum\lsu               % 6
\tutenum\beijing           % 7
\tutenum\berlin            % 8
\tutenum\bologna           % 9 
\tutenum\tata              % 10
\tutenum\ne                % 11
\tutenum\bucharest         % 12
\tutenum\budapest          % 13
\tutenum\mit               % 14 
\tutenum\debrecen          % 15
\tutenum\florence          % 16
\tutenum\cern              % 17 
\tutenum\wl                % 18 
\tutenum\geneva            % 19
\tutenum\hefei             % 20
\tutenum\seft              % 21
\tutenum\lausanne          % 22
\tutenum\lecce             % 23
\tutenum\lyon              % 24
\tutenum\madrid            % 25
\tutenum\milan             % 26
\tutenum\moscow            % 27
\tutenum\naples            % 27
\tutenum\cyprus            % 29
\tutenum\nymegen           % 30
\tutenum\caltech           % 31
\tutenum\perugia           % 32
\tutenum\cmu               % 33
\tutenum\prince            % 34
\tutenum\rome              % 35
\tutenum\peters            % 36
\tutenum\salerno           % 37
\tutenum\ucsd              % 38
\tutenum\santiago          % 39
\tutenum\sofia             % 40
\tutenum\korea             % 41
\tutenum\alabama           % 42
\tutenum\utrecht           % 43
\tutenum\purdue            % 44
\tutenum\psinst            % 45
\tutenum\zeuthen           % 46
\tutenum\eth               % 47
\tutenum\hamburg           % 48
\tutenum\taiwan            % 49
\tutenum\tsinghua          % 50
{
\parskip=0pt
\noindent
{\bf The L3 Collaboration:}
\ifx\selectfont\undefined%  old style font selection
 \baselineskip=10.8pt
 \baselineskip\baselinestretch\baselineskip
 \normalbaselineskip\baselineskip
 \ixpt
\else%                      new style font selection
 \fontsize{9}{10.8pt}\selectfont
\fi
\medskip
\tolerance=10000
\hbadness=5000
\raggedright
\hsize=162truemm\hoffset=0mm
\def\r{\rlap,}
\noindent

M.Acciarri\r\tute\milan\
P.Achard\r\tute\geneva\ 
O.Adriani\r\tute{\florence}\ 
M.Aguilar-Benitez\r\tute\madrid\ 
J.Alcaraz\r\tute\madrid\ 
G.Alemanni\r\tute\lausanne\
J.Allaby\r\tute\cern\
A.Aloisio\r\tute\naples\ 
M.G.Alviggi\r\tute\naples\
G.Ambrosi\r\tute\geneva\
% ...
\newpage
%\rule{\textwidth}{0.4pt}
\begin{list}{A}{\itemsep=0pt plus 0pt minus 0pt\parsep=0pt plus 0pt minus 0pt
                \topsep=0pt plus 0pt minus 0pt}
\item[\aachen]
 I. Physikalisches Institut, RWTH, D-52056 Aachen, FRG$^{\S}$\\
 III. Physikalisches Institut, RWTH, D-52056 Aachen, FRG$^{\S}$
\item[\nikhef] National Institute for High Energy Physics, NIKHEF, 
     and University of Amsterdam, NL-1009 DB Amsterdam, The Netherlands
\item[\mich] University of Michigan, Ann Arbor, MI 48109, USA
\item[\lapp] Laboratoire d'Annecy-le-Vieux de Physique des Particules, 
     LAPP,IN2P3-CNRS, BP 110, F-74941 Annecy-le-Vieux CEDEX, France
\item[\basel] Institute of Physics, University of Basel, CH-4056 Basel,
     Switzerland
\item[\lsu] Louisiana State University, Baton Rouge, LA 70803, USA
\item[\beijing] Institute of High Energy Physics, IHEP, 
  100039 Beijing, China$^{\triangle}$ 
\item[\berlin] Humboldt University, D-10099 Berlin, FRG$^{\S}$
\item[\bologna] University of Bologna and INFN-Sezione di Bologna, 
     I-40126 Bologna, Italy
\item[\tata] Tata Institute of Fundamental Research, Bombay 400 005, India
\item[\ne] Northeastern University, Boston, MA 02115, USA
\item[\bucharest] Institute of Atomic Physics and University of Bucharest,
     R-76900 Bucharest, Romania
\item[\budapest] Central Research Institute for Physics of the 
     Hungarian Academy of Sciences, H-1525 Budapest 114, Hungary$^{\ddag}$
\item[\mit] Massachusetts Institute of Technology, Cambridge, MA 02139, USA
\item[\debrecen] KLTE-ATOMKI, H-4010 Debrecen, Hungary$^\P$
\item[\florence] INFN Sezione di Firenze and University of Florence, 
     I-50125 Florence, Italy
\item[\cern] European Laboratory for Particle Physics, CERN, 
     CH-1211 Geneva 23, Switzerland
\item[\wl] World Laboratory, FBLJA  Project, CH-1211 Geneva 23, Switzerland
\item[\geneva] University of Geneva, CH-1211 Geneva 4, Switzerland
\item[\hefei] Chinese University of Science and Technology, USTC,
      Hefei, Anhui 230 029, China$^{\triangle}$
\item[\seft] SEFT, Research Institute for High Energy Physics, P.O. Box 9,
      SF-00014 Helsinki, Finland
\item[\lausanne] University of Lausanne, CH-1015 Lausanne, Switzerland
\item[\lecce] INFN-Sezione di Lecce and Universit\'a Degli Studi di Lecce,
     I-73100 Lecce, Italy
\item[\lyon] Institut de Physique Nucl\'eaire de Lyon, 
     IN2P3-CNRS,Universit\'e Claude Bernard, 
     F-69622 Villeurbanne, France
\item[\madrid] Centro de Investigaciones Energ{\'e}ticas, 
     Medioambientales y Tecnolog{\'\i}cas, CIEMAT, E-28040 Madrid,
     Spain${\flat}$ 
\item[\milan] INFN-Sezione di Milano, I-20133 Milan, Italy
\item[\moscow] Institute of Theoretical and Experimental Physics, ITEP, 
     Moscow, Russia
\item[\naples] INFN-Sezione di Napoli and University of Naples, 
     I-80125 Naples, Italy
\item[\cyprus] Department of Natural Sciences, University of Cyprus,
     Nicosia, Cyprus
\item[\nymegen] University of Nijmegen and NIKHEF, 
     NL-6525 ED Nijmegen, The Netherlands
\item[\caltech] California Institute of Technology, Pasadena, CA 91125, USA
\item[\perugia] INFN-Sezione di Perugia and Universit\'a Degli 
     Studi di Perugia, I-06100 Perugia, Italy   
\item[\cmu] Carnegie Mellon University, Pittsburgh, PA 15213, USA
\item[\prince] Princeton University, Princeton, NJ 08544, USA
\item[\rome] INFN-Sezione di Roma and University of Rome, ``La Sapienza",
     I-00185 Rome, Italy
\item[\peters] Nuclear Physics Institute, St. Petersburg, Russia
\item[\salerno] University and INFN, Salerno, I-84100 Salerno, Italy
\item[\ucsd] University of California, San Diego, CA 92093, USA
\item[\santiago] Dept. de Fisica de Particulas Elementales, Univ. de Santiago,
     E-15706 Santiago de Compostela, Spain
\item[\sofia] Bulgarian Academy of Sciences, Central Lab.~of 
     Mechatronics and Instrumentation, BU-1113 Sofia, Bulgaria
\item[\korea] Center for High Energy Physics, Adv.~Inst.~of Sciences
     and Technology, 305-701 Taejon,~Republic~of~{Korea}
\item[\alabama] University of Alabama, Tuscaloosa, AL 35486, USA
\item[\utrecht] Utrecht University and NIKHEF, NL-3584 CB Utrecht, 
     The Netherlands
\item[\purdue] Purdue University, West Lafayette, IN 47907, USA
\item[\psinst] Paul Scherrer Institut, PSI, CH-5232 Villigen, Switzerland
\item[\zeuthen] DESY, D-15738 Zeuthen, 
     FRG
\item[\eth] Eidgen\"ossische Technische Hochschule, ETH Z\"urich,
     CH-8093 Z\"urich, Switzerland
\item[\hamburg] University of Hamburg, D-22761 Hamburg, FRG
\item[\taiwan] National Central University, Chung-Li, Taiwan, China
\item[\tsinghua] Department of Physics, National Tsing Hua University,
      Taiwan, China
\item[\S]  Supported by the German Bundesministerium 
        f\"ur Bildung, Wissenschaft, Forschung und Technologie
\item[\ddag] Supported by the Hungarian OTKA fund under contract
numbers T019181, F023259 and T024011.
\item[\P] Also supported by the Hungarian OTKA fund under contract
  numbers T22238 and T026178.
\item[$\flat$] Supported also by the Comisi\'on Interministerial de Ciencia y 
        Tecnolog{\'\i}a.
\item[$\sharp$] Also supported by CONICET and Universidad Nacional de La Plata,
        CC 67, 1900 La Plata, Argentina.
\item[$\diamondsuit$] Also supported by Panjab University, Chandigarh-160014, 
        India.
\item[$\triangle$] Supported by the National Natural Science
  Foundation of China.
\item[\dag] Deceased.
\end{list}
}
\vfill

%%% Local Variables: 
%%% mode: latex
%%% TeX-master: t
%%% End:

%%% Local Variables: 
%%% mode: latex
%%% TeX-master: t
%%% TeX-master: t
%%% TeX-master: t
%%% End: 

%%% Local Variables: 
%%% mode: latex
%%% TeX-master: t
%%% End: 

%%% Local Variables: 
%%% mode: latex
%%% TeX-master: t
%%% End: 

%%% Local Variables: 
%%% mode: latex
%%% TeX-master: t
%%% End: 

\newpage


\begin{thebibliography}{99}

\bibitem{ref:QED}
S.~Tomonaga, Prog. Theor. Phys. {\bf 1} (1946) 27; J.~Schwinger, \PR {\bf 74}
  (1948) 1439; R.P.~Feynman, \PR {\bf 76} (1949) 749; \PR {\bf 76} (1949)
  769.

\bibitem{ref:codata}
CODATA Task Group, to be published in J. Phys. Chem. Reference Data {\bf 28},
  no. 6, 1999;\\ the results can also be found on WWW at
  \texttt{http://physics.nist.gov/constants}.

\bibitem{ref:running}
E.C.G.~St\"uckelberg and A.~Petermann, Helv. Phys. Acta {\bf 26} (1953) 499;
  M.~Gell-Mann and F.~Low, \PR {\bf 95} (1954) 1300; N.N.~Bogoliubov and
  D.V.~Shirkov, Dokl. AN SSSR 103 (1955) 203.

\bibitem{ref:amz1}
S.~Eidelman and F.~Jegerlehner,
  Z. Phys. {\bf C 67}  (1995) 585.

\bibitem{ref:burkhardt}
H.~Burkhardt and B.~Pietrzyk,
  Phys. Lett. {\bf B 356}  (1995) 398.

\bibitem{ref:steinhauser}
M.~Steinhauser,
  Phys. Lett. {\bf B 429}  (1998) 158.

\bibitem{ref:swartz}
M.L.~Swartz,
  Phys. Rev. {\bf D 53}  (1996) 5268;
A.D.~Martin and D.~Zeppenfeld,
  Phys. Lett. {\bf B 345}  (1994) 558.

\bibitem{ref:davier}
M.~Davier and A.~H\"ocker,
  Phys. Lett. {\bf B 419}  (1998) 419;
J.H.~K\"uhn and M.~Steinhauser,
  Phys. Lett. {\bf B 437}  (1998) 425;
R.~Alemany, M.~Davier and A.~H\"ocker,
  E. Phys. J. {\bf C 2}  (1998);
J.Erler,
  Phys. Rev. {\bf D 59}  (1999) 054008.

\bibitem{ref:adler}
S.~Eidelman, F.~Jegerlehner, A.L.~Kataev and O.~Veretin,
  Phys. Lett. {\bf B 454}  (1999) 369.

\bibitem{ref:topaz}
TOPAZ Collab., I.~Levine \etal,
  Phys. Rev. Lett. {\bf 78}  (1997) 424.

\bibitem{ref:venus}
VENUS Collab., S.~Odaka \etal,
  Phys. Rev. Lett. {\bf 81}  (1998) 2428.

\bibitem{L3-SLUM}
I.C.\ Brock \etal,
  Nucl. Inst. Meth. {\bf A 381}  (1996) 236.

\bibitem{ref:geant}
The L3 detector simulation is based on GEANT Version 3.15.\\ R.~Brun \etal,
  {\em GEANT 3}, CERN-DD/EE/84-1 (Revised), 1987.

\bibitem{ref:bhlumi}
BHLUMI version 4.04 is used. \\ S.~Jadach \etal, \CPC {\bf 70} (1992) 305; \PL
  {\bf B 353} (1995) 349; \PL {\bf B 353} (1995) 362; \CPC {\bf 102} (1997)
  229.

\bibitem{bhlumi-new}
B.F.L.~Ward, S.~Jadach, M.~Melles and S.A.~Yost,
  Phys. Lett. {\bf B 450}  (1999) 262.

\bibitem{l3-00}
L3 Collab., B. Adeva \etal,
  Nucl. Inst. Meth. {\bf A 289}  (1990) 35.

\bibitem{BHWIDE}
S.~Jadach, W.~Placzek and B.F.L.~Ward,
  Phys. Lett. {\bf B 390}  (1997) 298.

\bibitem{Placzek}
W.~Placzek, private communication.

\end{thebibliography}
\end{document}